\documentclass[12pt]{article}\usepackage[]{graphicx}\usepackage[]{xcolor}
\makeatletter
\def\maxwidth{ %
  \ifdim\Gin@nat@width>\linewidth
    \linewidth
  \else
    \Gin@nat@width
  \fi
}
\makeatother

\definecolor{fgcolor}{rgb}{0.345, 0.345, 0.345}

\usepackage{framed}
\makeatletter
 {\par\unskip\endMakeFramed%
 \at@end@of@kframe}
\makeatother

\definecolor{shadecolor}{rgb}{.97, .97, .97}
\definecolor{messagecolor}{rgb}{0, 0, 0}
\definecolor{warningcolor}{rgb}{1, 0, 1}
\definecolor{errorcolor}{rgb}{1, 0, 0}
\newenvironment{knitrout}{}{} % an empty environment to be redefined in TeX

\usepackage{alltt}
\usepackage{amsmath,amssymb,amsthm,hyperref,graphicx,natbib,geometry}
\usepackage{authblk}

\newcommand{\Exp}[2][]{{\rm E}_{#1}[ #2 ]  }
\newcommand{\Var}[1]{{\rm Var}[ \ensuremath{ #1 } ]  }

\IfFileExists{upquote.sty}{\usepackage{upquote}}{}
\begin{document}

\title{Evaluation of the linear mixing model in fluorescence spectroscopy}

\author[1]{Peter Hoff} 
\author[2]{Christopher Osburn}
\affil[1]{Department of Statistical Science, Duke University} 
\affil[2]{Department of Marine, Earth, and Atmospheric Sciences, North Carolina State University}

\maketitle

\begin{abstract} 
Analyses of spectral data often assume a \emph{linear mixing hypothesis}, 
which states that the spectrum of a mixed substance is approximately the 
mixture of the individual spectra of its constituent parts. 
We evaluate this hypothesis in the context of dissolved organic matter (DOM) fluorescence spectroscopy for endmember abundance recovery from mixtures of three different DOM endmembers. We quantify two key sources of experimental variation, and statistically evaluate the linear mixing hypotheses in the context of this variation. We find that there is not strong statistical evidence against this hypothesis for high-fluorescence readings, and that true abundances of high-fluorescence endmembers are accurately 
recovered from the excitation-emission fluorescence spectra of 
mixed samples using linear methods. However, abundances of a low-fluorescence endmember are less well-estimated, in that the abundance coefficient estimates exhibit a high degree of variability across replicate experiments. 
\end{abstract} \paragraph{Summary:} 
The impact of dissolved organic matter (DOM) on aquatic ecosystems is not always concentration-dependent: its sources matter. While the composition of DOM has been studied using sophisticated and cost-intensive chemical techniques, the light emitted by DOM, its fluorescence, is a rapid and cheap measurement technique that can fingerprint DOM’s sources in natural waters. However, fluorescence can only unravel a mixture of sources if the fluorescence of the mixture is approximately the same as a mixture of the source fluorescence. We evaluate this paradigm and confirm it holds for a range of DOM fluorescence profiles exhibited by most natural waters. However, we emphasize the need to make replicate measurements to confirm the reliability of the estimates made using linear unmixing methods.   

\section{Introduction}  
\subsection{Scientific Background} 
Dissolved organic matter (DOM) in natural waters is a heterogeneous mixture of hundreds to thousands of individual compounds of varying molecular complexity and reactivity. DOM forms the basis of the microbial loop in natural waters and constitutes a major component of Earth’s carbon cycle, while also influencing water quality in lakes, streams, groundwater, rivers, and estuaries and coastal waters. Along the hydrologic networks of the Earth’s surface connecting land to ocean, DOM originates from a variety of sources such as natural organic matter stored in surface plant litter and soils, produced within natural waters via primary and secondary production, and exchanged between adjacent ecosystems. Further, DOM is generated in the built environment and can be discharged into natural waters as domestic and industrial sewage as well as runoff from human modifications to the landscape from urbanization and agriculture (e.g.,  
\citet{wilson2009effects,bhattacharya2020spatial}). 

Unraveling the mixture of DOM sources is important because DOM’s biogeochemical reactivity reflects its composition (i.e., quality) as much as its concentration (quantity). DOM’s reactivity in natural waters is important because the degree to which it can be utilized by microorganisms can influence water quality – potentially contributing to eutrophication, hypoxia, and acidification. For researchers and practitioners alike, understanding the reactivity of DOM by characterizing its quality and then apportioning its sources in a mixture is important. 

Several approaches have been utilized to characterize DOM’s quality, and chief among them is its light absorbing and fluorescing properties, collectively termed the chromophoric portion of DOM – CDOM. These spectral properties of DOM can be measured rapidly (within a few hours of collection) on cost-effective instruments.  While they do not elucidate the molecular composition specifically, CDOM 
absorbance and fluorescence do provide a great deal of information about its sources \citep{fellman2010fluorescence,stedmon2015optical}. 

Across the land to ocean aquatic continuum (LOAC), DOM can be thought of as a mixture of several sources. Thus, with prior knowledge of the possible sources to a particular downstream location in a river or to receiving waters such as lakes and estuaries, it should be possible to unravel the components of the mixture. For example, endmember mixing models assume that runoff in a catchment is a mixture of unique sources whose contributions can be determined using tracers that exhibit conservative behavior, even if they integrate processes occurring in a watershed \citep{inamdar2011use,larsen2015fluorescence}.
Prior work with CDOM fluorescence as a conservative tracer of DOM sources has utilized parallel factor analysis (PARAFAC) decomposition 
\citep{bro1997parafac,osburn2016predicting} 
 and least squares regression 
\citep{goldman2012applications},
and endmember mixing analysis (EMMA) based on mass balance  
\citep{derrien2018estimation,lee2020comparing}. 
Recent work by \citet{bryan_hoff_osburn_2023,bryan_hoff_osburn_2024}  provides a simple least squares regression approach (``Regress-Then-Sum'', RTS) that add to the methodology available for DOM source unmixing. Uncertainty in the measurement of CDOM fluorescence has been addressed 
\citep{murphy2010measurement,korak2014critical}.
Implicit in these unmixing methods is the linear mixing hypothesis, which 
assumes 
that the fluorescence profile 
of a mixed sample resembles the mixed profile of its source components. This concept constitutes the rationale of the mass balance approach.   
In theory, a DOM sample taken along the LOAC is a mixture of the sources upstream to it. The mass balance approach requires, for example, that in a river network the fractions of the upstream endmembers in a downstream sample must sum to 1. However, each source and the resulting mixture carry some environmental history including microbial or photodegradation, adsorption or desorption onto or from particles, binding to metals, among others \citep{lee2018new}. Thus a mixed water sample will be a mixture of these upstream sources and the processing these upstream sources experience prior to sample collection. (In tidal systems, we may also consider ocean sources.) We may also represent source categories by measuring fluorescence on representative samples collected from a particular source and incorporating its environmental history. In principle, this means we can define a collection of fluorescence observations measured within source categories (i.e., a dictionary). With this framework in mind, we can now formalize the source mixing problem based on fluorescence.
It can be expected that DOM samples with substantial measurable signal will follow this assumption because the magnitude of experimental variation will be  much lower than the signal itself.
 However, a key question to answer is if the experimental variation for samples with fluorescence signals near the limit of detection is small enough make the linear mixing hypothesis useful in practice. This is important to establish, in particular when mixing problems must resolve very small individual contributions of sources (e.g., a few percent). 
Thus, to evaluate the utility of 
the linear mixing hypothesis as applied to CDOM fluorescence
there is a need to 
identify 
the range of fluorescence signals for which the hypothesis is likely to hold
and be practically useful for reliable estimation of mixing proportions.

\subsection{Statistical Background}  

As described below in Section 2, CDOM profiles of water samples are obtained 
by measuring fluorescence 
at combinations of multiple excitation and emission frequencies. 
These fluorescence measurements are generally arranged to form 
an excitation-emission matrix, or EEM. 
Each excitation/emission pair in the EEM is referred to as a \emph{pixel}. 
For mathematical and statistical calculations, it is often useful 
to rearrange the entries of an EEM as a vector, 
which we refer to as the 
 \emph{vectorization} of the EEM. 

Let $\mu$ be a vectorization of a 
hypothetical EEM from a mixed water sample 
that is measured with no replication or measurement error, 
and similarly define $\theta^k,k=1,\ldots,s$ as noiseless vectorized EEMs 
of  potential endmembers of the sample. If the abundances of the endmembers in the 
mixed sample are given by $b_1,\ldots, b_s$, then the linear mixing 
hypothesis states that 
\[
  \mu \approx  \Theta b  = \theta^1 b_1 + \cdots  + \theta^s b_s, 
\] 
where $\Theta$ is a matrix with columns $\theta_k, k =1,\ldots,s$, $b$ is a vector with elements $b_1,\ldots,b_s$, and $\Theta b$  is the matrix-vector product of $\Theta$ and $b$. 
Deviations from the linear mixing hypothesis 
could be evaluated, pixel by pixel, 
by examining the elements of the vector $ \mu - \Theta b$. 

Due to variation in experimental procedures 
and noise from measurement devices, the exact values of 
$\mu$ and $\theta^1,\ldots, \theta^s$ are unavailable. 
Instead, 
we have noisy measurements of these vectors, from which we 
obtain estimates $\hat\mu$ and $\hat\theta^1,\ldots, \hat\theta^s$. 
The linear mixing hypothesis can be evaluated by comparing 
$\hat\mu$ to $\hat\Theta b$, but even if the hypothesis is true 
we do not expect these quantities to be exactly equal, because 
$\hat\mu$ and $\hat\Theta$ are not exactly equal to 
$\mu$ and $\Theta$ due to experimental noise and variation. 
To evaluate the linear mixing hypothesis using $\hat\mu$ and $\hat\Theta b$, 
we need to compare the values of their difference to the range of plausible values, accounting for experimental variation. 

\subsection{Outline of This Study}  
In Section 2 
we describe an 
experiment conducted to quantify the sources of experimental variation 
in fluorescence measurements of several mixed water samples obtained from three known sources of DOM in the Neuse River basin, eastern North Carolina: groundwater, streamwater, and wastewater. These sources represent a gradient of CDOM concentrations common in natural waters of the LOAC 
\citep{massicotte2017global}. 
From our experimental results, 
in Section 3 
we identify and quantify
two primary sources of experimental variation, which we refer to as procedural variation and measurement variation. The degree of procedural variation among the water samples is fairly consistent, but the degree of measurement variation 
appears to be signal-dependent. Specifically, 
our groundwater samples with very low amounts of fluorescence near the limit of detection exhibit a much lower signal-to-noise ratio than the other samples.
Using estimates of these sources of variation, we statistically evaluate the linear mixing 
hypothesis in Section 4, and find little evidence of violations of 
the hypothesis for most combinations of excitation and emission wavelengths, especially for pixels with strong emission signals. 
We also evaluate the precision of the linear mixing model for 
endmember abundance estimation, 
by comparing 
the true endmember abundances in a mixed water sample to estimated 
abundance coefficients. For the water samples analyzed in this study, 
the abundances of high-fluorescence endmembers are precisely estimated, 
whereas abundance estimates for the low-fluorescence groundwater endmember 
are less precise. 
A discussion of results and recommendations follow in Sections 5 and 6. 

\section{Materials and Procedures}

We evaluated the linear mixing hypothesis using mixtures of three endmember water samples, including a groundwater sample, a streamwater sample and a wastewater sample. These samples were collected in late April and early May of 2024. The groundwater sample was collected from a residential well in Wake County, NC. The streamwater sample was collected from a walking bridge over Rocky Branch, a stream located on NC State’s campus. The wastewater sample was collected as a 24-hour composite sample of domestic and industrial sewage in the influent entering the Corpening Creek Wastewater Treatment Plant in Marion, NC.  All three samples were collected in detergent-cleaned brown HDPE bottles and stored at 4 $^\circ$C in the dark until transported to the laboratory. Samples were then filtered through pre-combusted (450 $^\circ$C for six hours) Cytiva Whatman glass fiber filters (GFF; 0.7 $\mu$m mesh size) into detergent-cleaned polycarbonate bottles and stored in a refrigerator until measurement. The wastewater was diluted (1:2 vol/vol) with ultrapure water (MilliQ, 18.2 M$\Omega$ resistivity) prior to filtration. Dilution was accounted for when fluorescence was corrected as described below. 

\begin{table}
\centering
\begin{tabular}{l||c|c|c||} 
\hline 
Sample&  \% Groundwater& \% Streamwater& \% Wastewater\\  \hline
$s_1$  & 1.00& 0.00& 0.00  \\ 
$s_2$  & 0.00& 1.00& 0.00  \\ 
$s_3$  & 0.00& 0.00& 1.00  \\ \hline
$m_1$ & 0.00& 0.50& 0.50 \\ 
$m_2$ & 0.50& 0.50& 0.00\\  
$m_3$ & 0.50& 0.00& 0.50 \\
$m_4$ & 0.25& 0.25& 0.50 \\
$m_5$ & 0.25& 0.50& 0.25 \\
$m_6$ & 0.50& 0.25& 0.25 \\
$m_7$ & 0.33& 0.33& 0.33  \\ \hline 
\end{tabular} 
\caption{Mixture weights (abundances) of endmembers of the ten water samples.}
\label{tab:mixprop} 
\end{table} 

The three different source water samples, or endmembers, were mixed in various proportions to obtain seven different mixed water 
samples (Table \ref{tab:mixprop}). 
These mixtures represent examples of problems where we wish to determine the relative proportions of a mixture. For example, contamination of wastewater in streamwater or the relative exchange between groundwater seepage discharging into a stream. For each of the $s = 3$ source samples and $m = 7$ mixed samples, $n = 3$ replicate aliquots (subsamples) were obtained and scanned. For all dilutions and sample transfers, a laboratory pipette (Gilson Pipetteman) with polypropylene tips were used. The pipette tips were rinsed with Milli-Q water prior to use. Measurement of Milli-Q water transferred with the pipette revealed no contamination of absorbance or fluorescence by the tips.  

Absorbance of each aliquot was scanned first from 200 to 800 nm in a 1 cm quartz cell (Starna) using a Varian Cary 300UV spectrophotometer. Milli-Q water was scanned separately and used as a blank. CDOM absorption values were computed as $a_\lambda = (A_{\lambda,\text{sample}} - A_{\lambda,\text{blank}})/L$, where  $a$ is the Napierian absorption coefficient, $A$ is the unitless absorbance measured on the spectrophotometer, $\lambda$ is the wavelength in nm, and $L$ is the pathlength of the quartz cell, in meters.  The mean and standard deviation of the CDOM absorption at 350 nm, a common expression of CDOM concentration in natural waters is given in Table \ref{tab:cdomabs}. These values fall within the large range of CDOM values for natural waters across the LOAC 
\citep{massicotte2017global}. 
CDOM fluorescence is measured as emission (Em) intensity over a wavelength range of constant interval at a fixed excitation (Ex) wavelength, creating multiple emission spectra, which get arranged to form an excitation-emission matrix (EEM) as described in Section 1.2. EEMs are most commonly viewed as contour plots or ``landscapes'' that visualize the strength of emission intensity in different regions of EEM-space.
Fluorescence emission from 300 to 600 nm at 2 nm increments was then measured at excitation wavelengths from 240 to 450 nm at 5 nm increments using a Varian Eclipse spectrophotometer. This procedure resulted in a $151 \times 43$-dimensional EEM for each of the 30 = $n \times (s + m)$ subsamples, representing fluorescence intensities at 151 emission wavelengths and 43 excitation wavelengths. For each EEM, the Milli-Q blank was subtracted, and then corrections applied for variability in excitation lamp spectra and detector response and inner filter effects
\citep{kothwala2013inner}.
Emission was normalized to the instrument’s water Raman signal and then calibrated to quinine sulfate units where 1 QSU = 1 $\mu$g/L quinine sulfate dissolved in 0.1 M sulfuric acid 
\citep{gilchrist2014optical}. 
Common absorbance and fluorescence properties of each source and mixture, including all replicates, are presented in the Supporting Information in Table S1.
Representative contour plots of the source and mixture EEMs are also provided in the Supporting Information. 

\begin{table}
    \centering
    \begin{tabular}{c||cc||} 
\hline
         Sample&  Mean&  Standard deviation\\ \hline
         Groundwater&  0.73&  0.44\\
         Streamwater&  7.92&  0.41\\
         Wastewater&  16.20&  0.12 \\ \hline
    \end{tabular}
    \caption{Mean and standard deviation of CDOM absorption values at 350 nm (units of inverse meters, $m^{-1}$) for the three samples used in this study.}
    \label{tab:cdomabs}
\end{table}

The physically meaningful entries of an EEM correspond to the $p = 5065$ 
excitation/emission pairs, or pixels, 
for which the emission wavelength is longer than the excitation wavelength. 
The $p$-dimensional vector of fluorescence intensities of an EEM 
at these pixels is the vectorization of the EEM. 
The statistical analyses that follow are based on the vectorizations of these 30 EEMs, and so the dataset consists of 30 $p$-dimensional vectorized EEMs, three for each of the ten water samples listed in Table \ref{tab:mixprop}. All analyses were performed using R Statistical Software 
\citep[v4.4.1]{rcitation}.  
Data and replication code for the numerical results in this article are 
available at 
\url{http://www.stat.duke.edu/~pdh10/Code/hoff_osburn_2024}

\section{Assessment of Experimental Variation} 

\subsection{Procedural Variation} 
First we investigate the variation across the $n=3$ replicate EEMs for each 
of the 10 water samples. 
Let $y_i$ be the $p$-dimensional 
vectorized EEM of the $i$th replicate from one of the mixed water samples, 
and let 
 $x_i$ be the 
vectorized EEM of the $i$th replicate from one of the 
endmember water samples. 
We expect replicate EEMs from  a common water sample to resemble each 
other, and thus to also resemble their within-sample average. 
Notationally, this means we expect that $y_i \approx \bar y$ and $x_i \approx \bar x$ 
for $i=1,\ldots, n$, where for example
$\bar x = \sum_i x_i/n$ is the $p$-dimensional vector 
obtained by averaging the replicate values at each pixel (i.e., 
$\bar x$ is the ``average EEM'' of an endmember). 

\begin{figure}[ht]

\begin{knitrout}\footnotesize
\definecolor{shadecolor}{rgb}{0.969, 0.969, 0.969}\color{fgcolor}

{\centering \includegraphics[width=1\linewidth]{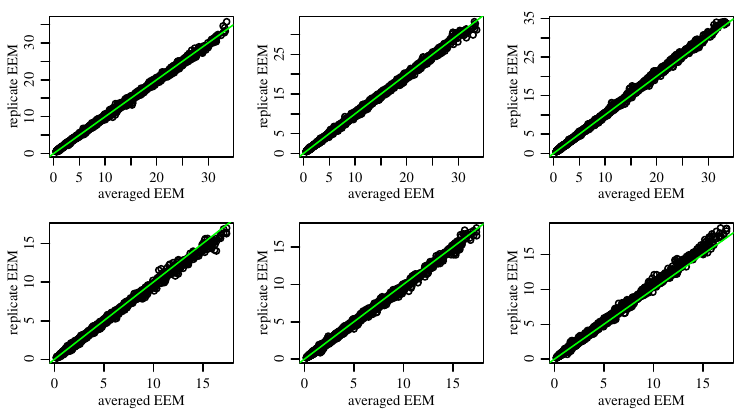} 

}

\end{knitrout}

\caption{Replicate versus average EEMs for two mixed water samples. The top row is mixture $m_1$ (streamwater and wastewater), the bottom row is mixture $m_2$ (streamwater and groundwater). }  
\label{fig:scatterplots}  
\end{figure}

This approximation is examined empirically for two 
of the mixture samples ($m_1$ and $m_2$) in Figure \ref{fig:scatterplots}. 
Mixture $m_1$ was 50:50 (vol/vol) of the streamwater and the wastewater, respectively. Mixture $m_2$ was 50:50 (vol/vol) of the streamwater and the groundwater. 
The figure shows a strong linear correlation between each replicate EEM
and the average EEM for its water sample. However, notice that the fluorescence values in several 
of the scatterplots are slightly but consistently above or below the green 45 degree line, indicating that the pixels of one replicate can have values systematically 
higher or systematically lower than those of another from the same water sample. 

A more comprehensive evaluation of this phenomenon is given in Table 
\ref{tab:expVar}. 
The first three columns are the linear correlations between 
each average EEM and the replicate EEMs that make up the average. 
For example, the first three columns give for each row 
the across-pixel sample correlations $\hat\rho_1,\hat\rho_2, \hat \rho_3$  between $\bar y$ and 
$y_1, y_2, y_3$ respectively  (or between 
$\bar x$ and $x_1,x_2,x_3$ for 
an endmember water sample). 
The correlation coefficients are all nearly 1, except for 
the replicates from the groundwater sample $s_1$.

\begin{table}[ht]
\centering

\begin{tabular}{l||ccc|ccc|c||ccc|ccc|c||ccc|ccc|c||ccc|ccc|c||ccc|ccc|c||ccc|ccc|c||ccc|ccc|c}
\hline
  & $\hat \rho_1$ & $\hat \rho_2$ & $\hat\rho_3$ & $\hat a_1$ & $\hat a_2$ & $\hat a_3$ & $\hat \sigma_a$\\
\hline
$s_1$ & 0.903 & 0.908 & 0.882 & 1.015 & 0.955 & 1.026 & 0.038\\
$s_2$ & 0.999 & 0.999 & 0.999 & 0.991 & 0.989 & 1.020 & 0.017\\
$s_3$ & 1.000 & 1.000 & 1.000 & 1.034 & 0.973 & 0.992 & 0.031\\
\hline
$m_1$ & 0.999 & 0.999 & 0.999 & 1.016 & 0.983 & 1.001 & 0.016\\
$m_2$ & 0.998 & 0.998 & 0.999 & 0.966 & 0.970 & 1.064 & 0.056\\
$m_3$ & 0.999 & 0.999 & 0.999 & 0.966 & 1.042 & 0.992 & 0.039\\
$m_4$ & 0.999 & 0.999 & 0.999 & 0.990 & 1.026 & 0.984 & 0.023\\
$m_5$ & 0.999 & 0.999 & 0.999 & 0.941 & 1.039 & 1.020 & 0.052\\
$m_6$ & 0.999 & 0.999 & 0.999 & 0.942 & 1.049 & 1.010 & 0.054\\
$m_7$ & 0.999 & 0.999 & 0.999 & 0.942 & 1.029 & 1.029 & 0.050\\
\hline
\end{tabular}

\caption{EEM-level multiplicative variation among water samples.  Rho ($\hat\rho$) is the Pearson's correlation coefficient, $\hat a$ is the across-pixel average ratio of a replicate to its mixture's average, and $\hat \sigma_a$ is the standard deviation of the average ratios.  }  
\label{tab:expVar} 
\end{table}

The second three columns of the table give the across-pixel average ratio of 
a replicate EEM to its corresponding average. 
For example, in each  row the coefficient $\hat a_i$ is 
given by  $\hat a_i= \sum_j (y_{i,j} / \bar y_j )/p$, 
where $i$ indexes replicates and  $j$ indexes pixels. 
Across the table, the $a$-values 
range from $0.941$ to $1.064$, and thus vary by 
as much as $6\%$ from 1. 
The final column of Table \ref{tab:expVar}
gives $\hat \sigma_a$, 
the sample standard deviation of these average ratios, that is, 
the sample standard deviation of the preceding three columns of the table. 
This number quantifies how much the pixel values in a replicate EEM may be 
systematically larger or smaller than those of the sample average 
EEM. 

A replicate EEM with a high $\hat a$-value indicates that, on average across pixels, its fluorescence values are higher than those of the other replicates from the same water sample. Possible explanations of this phenomenon include
variation in the amount of Milli-Q water added for dilution, 
variation in residual water in a sample container, 
and possibly other slight variations in experimental procedures  that involve use of the Gilson Pipetteman pipette. 
We refer to this variation 
as \emph{procedural variation}, and model this variation mathematically
 with the multiplicative  approximation  given by
\[
  y_i \approx a_i  \times \mu, 
\]
where
$\mu$ is a $p$-dimensional  perfectly-measured ``true'' EEM, 
 and $a_i$ is a scalar representing  how much the overall fluorescence of 
an observed EEM $y_i$ varies as a fraction of that of $\mu$.  
The values $a_1,\ldots, a_n$ can be viewed as multiplicative error 
terms that vary around the value of 1. 
An estimate of $\mu$ is given by  $\hat\mu = \bar y$, and as described above, 
we estimate each $a_i$ as $\hat a_i = \sum_{j} ( y_{i,j}/\hat \mu_j)/p$. 
Because the values of $\hat\sigma_a$ do not vary considerably across the 
samples, in the analyses that follow we use a 
single pooled estimate of $\hat\sigma_a \approx 0.04$, 
or equivalently, 
$\hat\sigma^2_a \approx 0.0016$. 

\subsection{Measurement Variation} 
\begin{figure}
\begin{knitrout}\footnotesize
\definecolor{shadecolor}{rgb}{0.969, 0.969, 0.969}\color{fgcolor}

{\centering \includegraphics[width=1\linewidth]{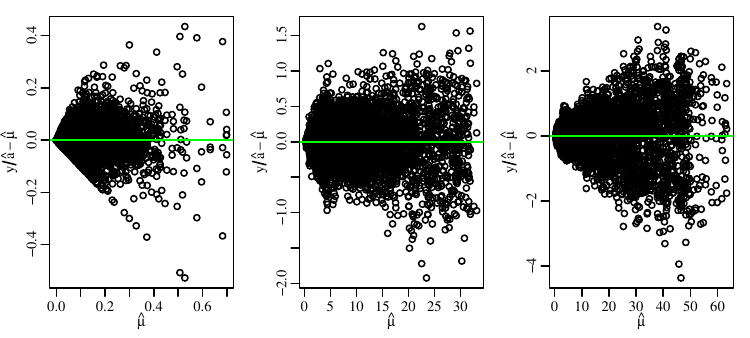} 

}

\end{knitrout}
\caption{Fitted versus residual plots for $s_1$, $s_2$ and $s_3$. } 
\label{fig:mrplot} 
\end{figure}

In the previous subsection we examined the approximation 
 $y_{i} \approx a_i \mu$
where $\mu$ is an unknown perfectly-measured  EEM vector for a given water 
sample, and $a_i$ is single number that scales the entire $\mu$-vector 
up or down. 
We estimated each $a_i$ by 
first estimating $\mu$ with the sample average vector 
$\hat\mu= \sum_i y_i/n$, and then letting 
$\hat a_i = \sum_j (y_{i,j} / \hat \mu_j)/p$.  Based on the above approximation 
we have $y_i/a_i \approx \mu$ and so we expect that
  $y_i /\hat a_i \approx \hat \mu$.
However, even after accounting for the across-replicate multiplicative 
variation 
represented by variation in $a_1,\ldots, a_n$, there is still 
pixel-to pixel variation. Some of this variation is displayed graphically 
in Figure \ref{fig:mrplot}, which for 
the three endmember water samples 
($s_1$, $s_2$ and $s_3$), plots the difference between each scale-adjusted EEM 
 $y_i/\hat a_i$ and the sample average $\hat \mu $, as a function 
of $\hat\mu$. 
These ``residuals'' of the scaled replicate EEMs  
vary around zero by an amount that increases with the magnitude of the 
average signal. In other words, the residual variance for high-fluorescence
pixels
is higher than that of low-fluorescence pixels. 
This kind of mean-variance relationship
is common for non-negative data types such as fluorescence data. 

\begin{figure}
\begin{knitrout}\footnotesize
\definecolor{shadecolor}{rgb}{0.969, 0.969, 0.969}\color{fgcolor}

{\centering \includegraphics[width=1\linewidth]{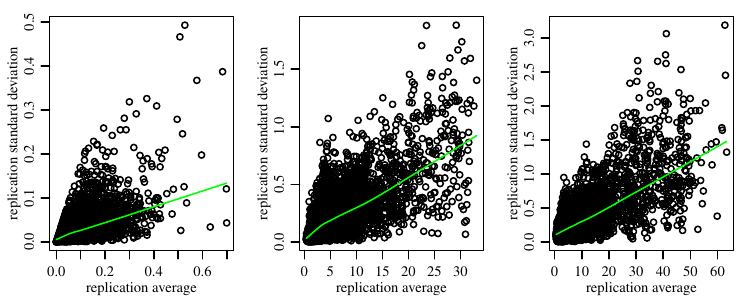} 

}

\end{knitrout}
\caption{Sample replication standard deviation versus sample replication average across pixels for water samples $s_1$, $s_2$ and $s_3$.} 
\label{fig:mvplot} 
\end{figure}

This mean-variance relationship can be quantified numerically by comparing 
the sample mean and sample standard deviation of the three replicate values at each 
pixel and for each endmember water sample. This relationship is shown graphically 
in Figure \ref{fig:mvplot}. Each point in each of the three plots   
 gives the sample mean and sample standard deviation of one of the 
$p$ pixels. The green curve in each figure is a smooth approximation of the 
functional relationship between the mean and the standard deviation. 
While noisy due to the small sample size ($n=3$ at each pixel), 
there is a clear positive mean-variance relationship that looks approximately
linear, on average across pixels. The pattern looks similar for the seven other water samples that were mixtures of the three endmembers. 

A linear relationship between the mean of a random variable and 
its standard deviation can be modeled with multiplicative error, rather 
than the usual additive error assumed in most statistical models. Specifically,
if we use the model 
\begin{equation} 
 y_{i,j}  = a_i \mu_j e_{i,j}  
\label{eqn:mmodel} 
\end{equation}
where $e_{i,j}$ has a mean of one and a standard deviation  $\sigma_e$, then 
the standard deviation of a scaled EEM measurement $y_{i,j}/a_i$
is 
\begin{align*} 
\text{SD}[ y_{i,j}/a_i ] & =  \text{SD}[\mu_j e_{,j} ] \\  
& =  \mu_j \text{SD}[ e_{,j} ]  = \mu_j \sigma_e,  
\end{align*} 
and so the standard deviations across pixels $j=1,\ldots,p$ will be linearly related 
to the means, with slope given by $\sigma_e$.

\begin{table}
\centering

\begin{tabular}{l||ccc|||ccc|}
\hline
  & $\hat \sigma_e$ & $\hat{\text{SNR}}$\\
\hline
$s_1$ & 0.155 & 6.358\\
\hline
$s_2$ & 0.002 & 279.789\\
\hline
$s_3$ & 0.003 & 230.323\\
\hline
$m_1$ & 0.002 & 264.960\\
\hline
$m_2$ & 0.004 & 188.193\\
\hline
$m_3$ & 0.005 & 154.957\\
\hline
$m_4$ & 0.003 & 234.333\\
\hline
$m_5$ & 0.002 & 251.058\\
\hline
$m_6$ & 0.004 & 194.279\\
\hline
$m_7$ & 0.003 & 229.615\\
\hline
\end{tabular}

\caption{Sample-specific estimates of $\sigma_e$ (first column) and SNR (second column). }
\label{tab:stab} 
\end{table}

Estimates of $\sigma_e$ can be obtained by computing the 
standard deviation of the multiplicative residuals $\hat e_{i,j}$, given by
\[
 \hat e_{i,j} = y_{i,j}/(\hat a_i \hat \mu_j ). 
\]
Separate estimates of $\sigma_e$  for each water sample are given in Table 
\ref{tab:stab}. The values are similar across all water samples except 
for $s_1$, the groundwater endmember sample, having a standard deviation 
nearly an order of magnitude larger than the others. 

The estimates of $\sigma_e$ can be combined with the estimate of $\sigma_a$
to obtain signal to noise ratios for each of samples. 
Based on the multiplicative model (\ref{eqn:mmodel}), the variance 
of $y_{i,j}$ is given by 
\begin{equation} 
\Var{ y_{i,j} } =  \mu_j^2 \times \left( (\sigma_a^2 +1)\times (\sigma_e^2+1) - 1 \right ), 
\label{eqn:mvrel}  
\end{equation} 
and so the signal-to-noise ratio (SNR) for a given water sample is 
\[
 \text{SNR} = \mu_j^2 / \Var{ y_{i,j} } =  \frac{1}{(\sigma_a^2 +1)\times (\sigma_e^2+1) - 1  }
\]
Estimates of the SNR for each water sample are given in the 
second column of Table \ref{tab:stab}, using 
the common estimate of  $\hat\sigma^2_a$ described in the previous 
subsection and the sample-specific estimates of $\sigma^2_e$
 from the first column of the table. Note the extremely low SNR of the 
groundwater sample, relative to those of the other two. 

\section{Evaluating the Linear Mixing Model} 

\subsection{Testing the Linear Mixing Hypothesis} 
Recall from the Introduction that the 
 linear mixing 
hypothesis states that for each pixel $j$, 
\[\mu_j \approx \theta_j \cdot  b  = \theta^1_j b_1 + \cdots + \theta^s_j b_s\]
where $\theta_j = ( \theta^1_j ,\ldots, \theta^s_j)$ is 
the vector of fluorescence values at pixel $j$ of the $s$ endmember EEMs.
In this subsection, we statistically test 
the hypotheses $H_j: \mu_j =  \theta_j \cdot  b$, separately 
for each $j=1,\ldots,p$ and each of the seven mixed water samples, using the following  multiplicative model developed in the previous section:
\begin{itemize}
\item $y_{i,j}  = a_i \mu_j e_{i,j}$ for $i=1,\ldots,n$;
\item  $a_1,\ldots, a_n$ are independent and identically distributed  (i.i.d.)\ random variables with mean 1 
and variance $\sigma_a^2$; 
\item  $e_{1,j},\ldots, e_{n,j}$ are i.i.d.\ random variables with mean 1 
and variance $\sigma_e^2$;  
\item the $a_i$'s and $\epsilon_{i,j}$'s are independent. 
\end{itemize}
We also assume the same model holds for the replicates 
$x_{1,j}^k,\ldots, x_{n,j}^k$ for each endmember $k=1,\ldots, s$. 

Let $\hat \theta_j = \bar x_j = ( \bar x_j^1 ,\ldots, \bar x_j^s)$ be 
the vector of average fluorescence values for each of the $s$ 
endmembers at pixel $j$, and similarly let 
$\hat\mu_j = \bar y_j = \sum_i y_{i,j} /n$ be the sample 
average for a mixed water sample at pixel $j$. 
Under the model described above, 
$\hat\theta_j$ is an unbiased estimate of $\theta_j$, 
and $\hat \mu_j$ is an unbiased estimate of $\mu_j$.  
Evidence against the hypothesis $\mu_j = \theta_j \cdot b$, 
or equivalently $\mu_j -\theta_j \cdot b=0$, 
may be quantified by how far the estimate 
$\hat \mu_j -\hat \theta_j \cdot b$ is from zero, relative to its 
standard deviation. Specifically, we evaluate $H_j$ with the 
 $z$-statistic 
\[
  z_j = \frac{ \hat\mu_j - \hat \theta_j \cdot b}{ \hat \sigma_j} 
\]
where $\hat\sigma_j$ is an estimate of the standard deviation of the 
numerator in the above expression, and is derived in the 
Supporting Information.  
Under the model assumptions, if the 
linear mixing hypothesis is correct then 
the numerator of 
$z_j$ should have a mean of zero and be approximately normally distributed 
by the central limit theorem. Dividing by an estimate of the standard 
deviation provides us with a test statistic $z_j$ that should be approximately 
normally distributed under the linear mixing hypothesis. 
Thus a statistical test of $H_j: \mu_j = \theta_j \cdot b$ may be performed 
by comparing $z_j$ to a standard normal distribution.

\begin{figure}
\begin{knitrout}\footnotesize
\definecolor{shadecolor}{rgb}{0.969, 0.969, 0.969}\color{fgcolor}

{\centering \includegraphics[width=1\linewidth]{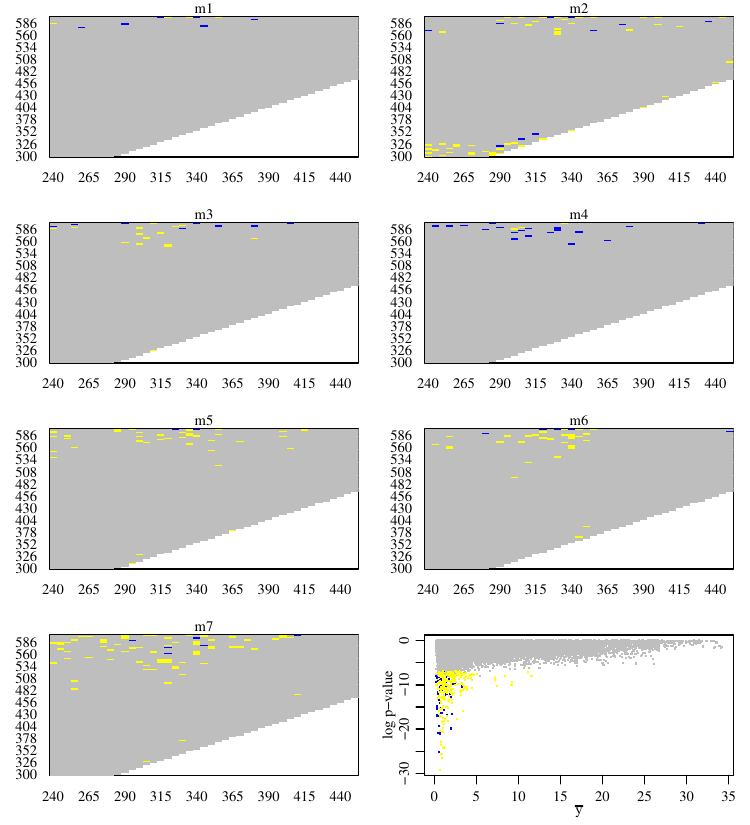} 

}

\end{knitrout}
\caption{Pixel-specific tests of the linear mixing hypothesis using the BH threshold with  $\text{FDR}=0.05$. Blue and yellow pixels indicate fluorescence of a mixed pixel as significantly lower and  higher than expected under linear mixing, respectively. 
The bottom-right panel plots the relationship 
between the average fluorescence at a pixel and the log $p$-value.
\label{fig:bhp}  
}
\end{figure}

For each of the $m$ mixed samples we computed a $z$-score 
$z_j$ and the associated $p$-value 
$p_j$ 
for each pixel $j=1,\ldots, p$. 
Significant deviations from the linear mixing hypothesis were 
evaluated using the Benjamini-Hochberg (BH) algorithm 
\citep{benjamini_hochberg_1995}
with a 
false discovery rate (FDR) of $\alpha=.05$ 
to account for the fact that we are evaluating $p=5065$ hypotheses 
for each mixed sample. 
Pixels having  $p$-values below the BH threshold are indicated 
in Figure \ref{fig:bhp}, with 
blue and yellow indicating 
pixels with lower and higher than expected fluorescence, 
respectively. 
Across samples, the number of such pixels 
was roughly $2\%$
or less of the total number of pixels. Most of the  the pixels 
that indicated some violation of the hypothesis had low-fluorescence 
measurements, although for 
most low-fluorescence pixels there is no strong 
evidence of the hypothesis being violated. 
 Moreover, the region of fluorescence where some violation of the hypothesis was indicated (emission wavelengths $\geq$500 nm) typically lacks DOM source information. 

The relationship between fluorescence and evidence against the 
linear mixing hypothesis is shown in the bottom-right panel 
of Figure \ref{fig:bhp}, 
which plots the $p$-value for each pixel on a log scale versus 
the sample average fluorescence at the pixel. 
In general, evidence against the hypothesis 
increases as fluorescence decreases. 
Even though most low-fluorescence pixels do not show strong evidence against linear mixing, 
these results suggest caution when using linear unmixing methods 
for abundance estimation of low-fluorescence endmembers. 
For example, nearly all of the pixels of our replicate
groundwater samples had fluorescence values less than 1 QSU, which 
is the range where approximately 
$40 \%$ of the violations 
of the linear mixing hypothesis are detected. 
We compared the results in Figure \ref{fig:bhp} that indicated significantly higher or lower fluorescence than expected under linear mixing with their corresponding EEMs (see Figure S2 in the Supporting Information).  Significantly higher or lower fluorescence was found over a wide range of excitation (240-415 nm) but restricted to emission $\geq$500 nm for most mixtures. Two exceptions are noted, for $m_2$ and $m_7$. In the case of $m_2$, overestimation was found in the peak T region of protein-like fluorescence 
\citep{coble2007marine}. 
For $m_7$, overestimation was found across a similar excitation range but at emission wavelengths $\geq$480 nm.  However, examination of the EEMs in Figure S2 clearly show the majority of fluorescence occurs outside of the regions noted above where significantly higher or lower fluorescence was estimated. This indicates that despite being significant, these regions of over- or underestimation carry low information and do not contribute to the fluorescence signal used to identify sources in mixtures.

\subsection{Experimental Variability of Abundance Estimates}
The previous results indicated that there is not strong evidence 
against the linear mixing hypothesis for the vast majority of pixels of an EEM,
especially pixels with high fluorescence intensities. 
However, estimation accuracy of 
endmember abundances relies  not 
just on the linear mixing hypothesis, but also on the 
strength of the fluorescence signal relative to experimental noise. 
As our groundwater sample exhibits both a low signal-to-noise ratio, and has pixel intensities in the range where the linear mixing hypothesis is more suspect, we expect more difficulty in obtaining 
reliable abundance estimates for this low-fluorescence endmember. 

Consider endmember abundance estimation for  a
mixed EEM $y$ 
using  noisy 
measured EEMs $x^1, \ldots, x^s$, one 
from each of the endmembers of which $y$ is assumed to be a mixture. 
Under the linear mixing hypothesis, the expectation of $y$ is 
\[
  \Exp{y} = \Theta b = \theta^1 b_1 + \cdots + \theta^s b_s
\]
where $b$ is the $s$-dimensional vector of abundances to be estimated
and $\theta^1,\ldots, \theta^s$ are hypothetical noiseless EEMs from 
each of the endmembers, so that $\Exp{ x^k} = \theta^k$ for $k=1,\ldots, s$. 
If  $\Theta$ were known then a simple estimate of $b$ would be 
the non-negative least squares estimate from the linear regression of 
$y$ on $\Theta$. As $\Theta$ is not observed, 
a simple alternative estimate $\hat b$ of $b$ 
is the non-negative least squares estimate of $y$ on 
the matrix $X$ having columns
$x^1,\ldots, x^s$. 

The estimate $\hat b$ of $b$  
will vary from experiment to experiment due to the  
variation of $X$ around $\Theta$ and of $y$ around $\Theta b$. 
In simple linear regression scenarios with independent, additive measurement
noise with variance $\sigma^2$, it is well-known that the 
variance of $\hat b$ 
is given by $\Var{ \hat b } = (X^\top X)^{-1}\sigma^2$. 
Roughly speaking, the variance of $\hat b_k$, the $k$th element of $\hat b$,
will be small if the magnitude of $x^k$, the $k$th column of $X$,  is large 
relative to $\sigma^2$, i.e.\ the signal to noise ratio is large. 
The situation for abundance estimation from fluorescence data is more complex, due to the multiple sources of  multiplicative variation and the fact that 
$x^1,\ldots,x^s$ are themselves measured with noise. 
However, we still 
expect that 
the precision of abundance estimates will be positively related to the signal
to noise ratio of the endmembers. 

While we lack a simple formula for the variance of $\hat b$, 
it can still  be assessed empirically if 
replicate EEMs are available, using the following resampling scheme 
which we illustrate 
in the context of the data from our study.
For each of our $m=7$ mixed samples, using non-negative least squares we regress one of the replicate mixed EEM vectors $y$ on the matrix $X$ consisting of one of the replicate groundwater EEMs, one 
of the replicate streamwater EEMs and one of the replicate wastewater EEMs. 
This generates a single estimate $\hat b$ of $b$. We repeat this procedure 
for each of the $3\times 3\times 3\times 3$ possible combinations 
of mixed and endmember EEMs, yielding 81 different estimates of $b$. 
The variation across these different estimates summarizes the effect 
of the experimental variability on the abundance parameter estimates.

\begin{figure}
\begin{knitrout}\footnotesize
\definecolor{shadecolor}{rgb}{0.969, 0.969, 0.969}\color{fgcolor}

{\centering \includegraphics[width=\maxwidth]{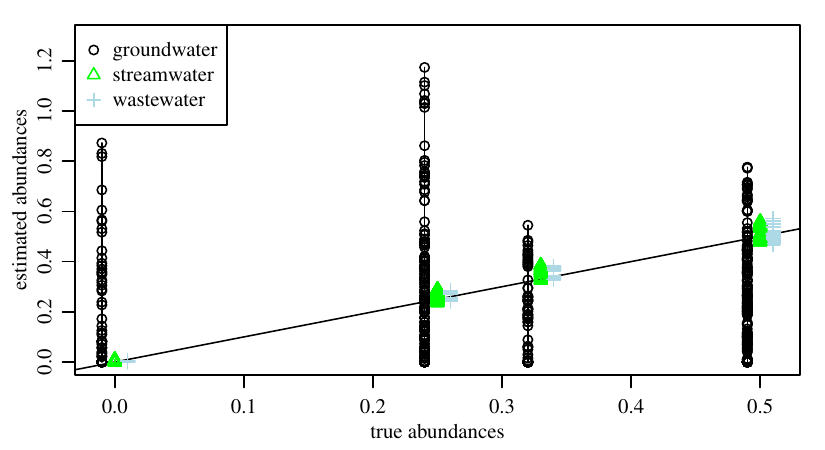} 

}

\end{knitrout}
\caption{NNLS abundance estimates. Coefficients for stream, waste and groundwater given in green, blue and black respectively.} 
\label{fig:estVar} 
\end{figure}

The results of this procedure are summarized graphically  in 
Figure \ref{fig:estVar}. Here, the parameter estimates 
 for the abundances 
of the groundwater, streamwater and wastewater 
endmembers are given in black, green and blue respectively. 
The estimates for 
the stream and wastewater endmembers are very close 
to the true mixing proportions, and have a very narrow range across the combinations of replicate EEMs, 
indicating a high degree of 
accuracy in these estimates. 
In contrast, the groundwater abundance estimates have a very large range and 
thus low accuracy.  The low 
signal to noise ratio of this endmember limits 
the precision to which its abundance in the mixture can be estimated. 

These results are summarized further in Table \ref{tab:estVarTab}. 
The columns give the true abundances $b_1$, $b_2$ and $b_3$ for 
each mixture, as well as the sample mean and standard deviation
of the 81 values for each of $\hat b_1$, $\hat b_2$ and $\hat b_3$. 
While the streamwater and wastewater estimates differ from the 
true mixing proportions slightly, they have low variance and are close enough 
for most applications. The groundwater estimates are not centered near the true mixing proportions, and the variance is high, indicating unreliability of 
abundance estimates for this endmember.

\begin{table}
\centering

\begin{tabular}{l|ccc|ccc|ccc|ccc|ccc|ccc|ccc|ccc|ccc|ccc|ccc|ccc|ccc|ccc|ccc|ccc|ccc|ccc|ccc|ccc|ccc|ccc|ccc|ccc|ccc|ccc|ccc}
\hline
\multicolumn{1}{c|}{ } & \multicolumn{3}{c|}{groundwater} & \multicolumn{3}{c|}{streamwater} & \multicolumn{3}{c}{wastewater} \\
\cline{2-4} \cline{5-7} \cline{8-10}
  & $b_1$ & mean($\hat b_1$) & sd($\hat b_1$) & $b_2$ & mean($\hat b_2$) & sd($\hat b_2$) & $b_3$ & mean($\hat b_3$) & sd($\hat b_3$)\\
\hline
$m_1$ & 0.00 & 0.153 & 0.228 & 0.50 & 0.496 & 0.009 & 0.50 & 0.491 & 0.014\\
$m_2$ & 0.50 & 0.386 & 0.197 & 0.50 & 0.517 & 0.022 & 0.00 & 0.003 & 0.001\\
$m_3$ & 0.50 & 0.169 & 0.224 & 0.00 & 0.004 & 0.004 & 0.50 & 0.537 & 0.028\\
$m_4$ & 0.25 & 0.337 & 0.370 & 0.25 & 0.250 & 0.007 & 0.50 & 0.497 & 0.017\\
$m_5$ & 0.25 & 0.227 & 0.193 & 0.50 & 0.531 & 0.027 & 0.25 & 0.268 & 0.013\\
$m_6$ & 0.50 & 0.198 & 0.162 & 0.25 & 0.272 & 0.014 & 0.25 & 0.268 & 0.014\\
$m_7$ & 0.33 & 0.158 & 0.166 & 0.33 & 0.358 & 0.020 & 0.33 & 0.361 & 0.019\\
\hline
\end{tabular}

\caption{Mean and standard deviation of resampled NNLS abundance estimates.}  
\label{tab:estVarTab}
\end{table}

\section{Discussion} 
A general class of 
 EMMA methods for estimating endmember abundance assume some version of a linear mixing hypothesis – that is, the fluorescence properties of a mixed sample resemble the mixture of properties of its endmembers. 
Our study provides experimental credibility to these approaches, 
in that statistical tests failed to reject the 
linear mixing hypothesis for most pixels of several mixed EEMs, 
and in particular, 
pixels with high fluorescence levels. 
For DOM fluorescence, most unmixing methods have used derived parameters from fluorescence such as discrete ``peak picking'' 
\citep{coble2007marine,goldman2012applications}, 
the fluorescence index (FI; 
\citet{mcknight2001spectrofluorometric}), 
and/or the biological index (BIX) and humification index (HIX) 
\citep{huguet2009properties}  
to distinguish among samples, in addition to decomposition methods such as parallel factor analysis (PARAFAC)
\citep{stedmon2003tracing}. 
Employing DOM fluorescence in endmember mixing analysis has typically involved such peak-picking or PARAFAC decomposition 
\citep{larsen2015fluorescence,osburn2016predicting}, 
and often in combination 
\citep{lee2020comparing}.  
As an alternative to using derived indices or PARAFAC components, 
\citet{bryan_hoff_osburn_2023} 
have shown how regressing the entire vectorized EEM of a mixed sample 
on the EEMs of its endmembers
can outperform other methods in terms of endmember abundance estimation. 
Our work provides support for the use of such direct linear estimation 
methods, as 
we were able to obtain 
highly accurate abundance estimates for high signal-to-noise endmembers 
using simple linear regression. 

This work also demonstrates that replication is critically important for evaluating the validity and variability of an unmixing method's estimated abundances. 
%Absent replication, the linear mixing hypothesis cannot be tested, and 
% accuracies of abundance parameter estimates are unknown. 
%Previous studies lacking replication  
%We have shown that linearity holds in an unmixing experiment with limited replication ($n=3$). However, triplicate samples are amenable to sample collection in many studies of natural waters, so our results are widely applicable. Despite the low replication, 
%this analysis was important because  
Variation in parameter estimates can arise due to a number of experimental 
factors. 
%In general, several aspects of fluorescence measurement can create uncertainty among replicates. 
For example, a bandpass resolution setting on a fluorometer greater than the emission increment can make it difficult to resolve peak emission wavelength, leading to variation among replicates \citep{korak2014critical}. 
High scan rate settings on fluorometers often create noisy spectra, further introducing variation. Our fluorescence measurements were taken with 5 nm bandpass settings and 2 nm emission increments at a scan rate of 2400 nm/min. These parameters were optimized for measuring large numbers of samples – a consideration important to many studies using fluorescence – so it was encouraging that results showed linear mixing held in the regions of fluorescence containing the most signal, and therefore the most important information, despite this scan speed. It is possible that with lower scan speeds the variation in the groundwater endmember could be reduced.  

In this work we have demonstrated how analysts can utilize replicates of discrete samples to measure experimental variation in their own laboratories. Such measures provide a means for statistical model evaluation 
(such as the linear mixing hypothesis) and 
assessments of estimation variability. In our study, 
we evaluated the linear mixing assumption over a range of fluorescence intensities commonly found in natural waters. 
We also showed that
estimates of stream and wastewater abundances are highly accurate, 
whereas those for groundwater are highly variable across 
replications. Absent replication and the resulting quantification 
of experimental variation, this assessment would be unavailable - 
the differential accuracy of the abundance 
estimates across endmembers would be unknown. 
%We cannot say anything about the environmental variability of these types of samples because we only utilized within-sample replication, evaluating LM compared to our laboratory’s experimental variation. We can say that for one analyst in our laboratory, from experiment to experiment, replicate to replicate, that the coefficients are reliable with a S/N of 6 or greater. 

\bibliographystyle{plainnat}
\bibliography{refs} 

\newpage 

\appendix

\renewcommand{\thefigure}{S\arabic{figure}}
\setcounter{figure}{0}
\renewcommand{\thetable}{S\arabic{table}}
\setcounter{table}{0}

\section*{Supporting Information} 

\subsection*{S.1 Absorbance and fluorescence parameters} 

\begin{table}[ht]  
\begin{centering} 
\begin{footnotesize} \begin{tabular}{|l|r|r|r|r|r|r|r|r|r|r|r|r|r|}
\hline
& & \multicolumn{3}{|l|}{$a_{254} \ (m^{-1})$}    & 
\multicolumn{3}{|l|}{$a_{350} \ (m^{-1})$}    & 
\multicolumn{3}{|l|}{$a_{440} \ (m^{-1})$}    & 
\multicolumn{3}{|l|}{$S_{300-650} \ (\mu m^{-1})$}     \\
\hline 
Sample & $n$ & mean & std & sem & mean & std & sem  & mean & std & sem  & mean & std & sem \\ \hline
$s_1$ & 3 & 1.400 & 0.590 & 0.340 & 0.73 & 0.44 & 0.25 & 0.37 & 0.19 & 0.11 & 16.80 & 7.16 & 4.13\\
\hline
$s_2$ & 3 & 28.980 & 0.530 & 0.310 & 7.92 & 0.41 & 0.24 & 1.88 & 0.13 & 0.08 & 15.58 & 0.20 & 0.12\\
\hline
$s_3$ & 3 & 56.570 & 0.100 & 0.060 & 8.10 & 0.12 & 0.07 & 3.19 & 0.07 & 0.04 & 15.35 & 0.39 & 0.23\\
\hline
$m_1$ & 3 & 43.510 & 1.800 & 1.040 & 8.19 & 0.96 & 0.55 & 2.72 & 0.30 & 0.17 & 15.03 & 0.47 & 0.27\\
\hline
$m_2$ & 3 & 15.450 & 0.620 & 0.360 & 4.13 & 0.78 & 0.45 & 1.22 & 0.11 & 0.07 & 15.35 & 0.36 & 0.21\\
\hline
$m_3$ & 3 & 29.560 & 0.590 & 0.340 & 5.04 & 0.45 & 0.26 & 1.87 & 0.12 & 0.07 & 14.52 & 0.32 & 0.19\\
\hline
$m_4$ & 3 & 36.050 & 0.290 & 0.170 & 6.66 & 0.10 & 0.06 & 2.24 & 0.13 & 0.08 & 14.97 & 0.40 & 0.23\\
\hline
$m_5$ & 3 & 29.030 & 0.810 & 0.460 & 6.24 & 0.43 & 0.25 & 1.94 & 0.19 & 0.11 & 15.19 & 0.40 & 0.23\\
\hline
$m_6$ & 3 & 21.970 & 1.180 & 0.680 & 4.46 & 0.77 & 0.44 & 1.48 & 0.10 & 0.06 & 14.92 & 0.10 & 0.06\\
\hline
$m_7$ & 3 & 29.790 & 1.150 & 0.660 & 5.75 & 0.71 & 0.41 & 1.87 & 0.16 & 0.09 & 15.14 & 0.26 & 0.15\\
\hline 
&  &  \multicolumn{3}{|l|}{$S_R$}    & 
\multicolumn{3}{|l|}{Fl}    & 
\multicolumn{3}{|l|}{BIX}    & 
\multicolumn{3}{|l|}{HIX}     \\  \hline
Sample & $n$ & mean & std & sem & mean & std & sem  & mean & std & sem  & mean & std & sem \\ 
\hline
$s_1$ & 3 & 0.813 & 0.926 & 0.535 & 1.38 & 0.74 & 0.42 & 1.77 & 0.80 & 0.46 & 3.88 & 3.56 & 2.06\\
\hline
$s_2$ & 3 & 0.803 & 0.001 & 0.001 & 1.35 & 0.03 & 0.02 & 0.75 & 0.02 & 0.01 & 5.18 & 0.12 & 0.07\\
\hline
$s_3$ & 3 & 2.979 & 0.074 & 0.043 & 1.70 & 0.04 & 0.02 & 0.84 & 0.05 & 0.03 & 0.46 & 0.01 & 0.01\\
\hline
$m_1$ & 3 & 1.947 & 0.028 & 0.016 & 1.56 & 0.08 & 0.04 & 0.79 & 0.00 & 0.00 & 1.01 & 0.00 & 0.00\\
\hline
$m_2$ & 3 & 0.859 & 0.048 & 0.028 & 1.34 & 0.04 & 0.02 & 0.74 & 0.03 & 0.01 & 5.11 & 0.05 & 0.03\\
\hline
$m_3$ & 3 & 2.841 & 0.158 & 0.091 & 1.74 & 0.30 & 0.18 & 0.89 & 0.04 & 0.02 & 0.46 & 0.01 & 0.01\\
\hline
$m_4$ & 3 & 2.257 & 0.060 & 0.035 & 1.58 & 0.16 & 0.09 & 0.83 & 0.05 & 0.03 & 0.76 & 0.03 & 0.02\\
\hline
$m_5$ & 3 & 1.559 & 0.074 & 0.042 & 1.50 & 0.13 & 0.07 & 0.83 & 0.05 & 0.03 & 1.44 & 0.02 & 0.01\\
\hline
$m_6$ & 3 & 1.872 & 0.004 & 0.002 & 1.48 & 0.20 & 0.12 & 0.83 & 0.01 & 0.01 & 1.03 & 0.00 & 0.00\\
\hline
$m_7$ & 3 & 1.896 & 0.088 & 0.051 & 1.52 & 0.10 & 0.06 & 0.79 & 0.03 & 0.02 & 1.02 & 0.02 & 0.01\\
\hline
\end{tabular}  
\end{footnotesize} 
\end{centering} 
\caption{Absorbance and fluorescence parameters for the samples used in this study.} 
\end{table}

\newpage

\subsection*{S.2 Example EEMs}

\begin{figure}[ht]
\begin{knitrout}\footnotesize
\definecolor{shadecolor}{rgb}{0.969, 0.969, 0.969}\color{fgcolor}

{\centering \includegraphics[width=5in]{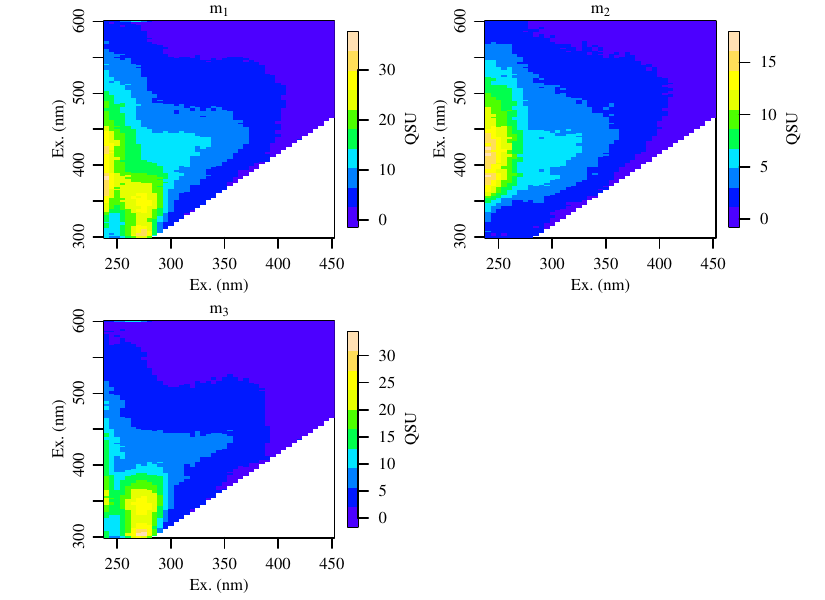} 

}

\end{knitrout}

\caption{One replicate EEM from each of the endmember water samples 
   $s_1$, $s_2$, $s_3$. }
\end{figure}

\begin{figure}
\begin{knitrout}\footnotesize
\definecolor{shadecolor}{rgb}{0.969, 0.969, 0.969}\color{fgcolor}

{\centering \includegraphics[width=5in]{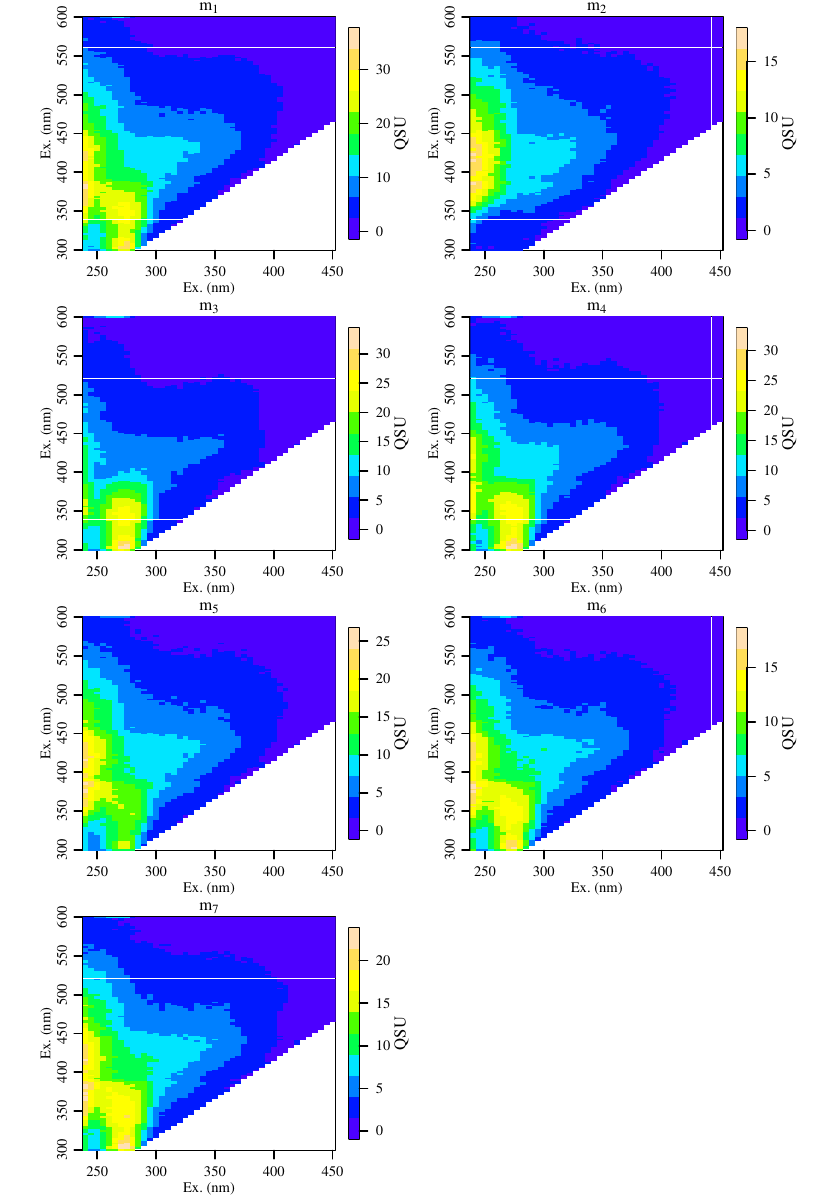} 

}

\end{knitrout}

\caption{One replicate EEM from each of the mixed  water samples 
   $m_1,\ldots,m_7$. }
\end{figure}

\newpage

\subsection*{S.3 Derivation of the linear mixing test statistic} 
We evaluate the hypothesis $H_j: \mu_j = \theta_j\cdot b$ with the 
test statistic 
$z_j = (\hat\mu_j - \hat \theta_j \cdot b)/\hat \sigma_j$, 
where  
\begin{itemize}
\item $\hat \mu_j = \bar y_j = \sum_{i=1}^n y_{i,j}/n$, the across-replicate average fluorescence at pixel $j$; 
\item $\hat \theta_j = ( \bar x_j^1 ,\ldots, \bar x_j^s)$, 
where $\bar x_j^k = \sum_{i=1}^n x_{i,j}^k/n$, the across-replicate average 
fluorescence at pixel $j$ for endmember $k$, $k=1,\ldots,s$.  
\item $\hat \sigma_j$ is an estimate of the standard deviation of  
$\hat \mu_j - \hat \theta_j \cdot b$. 
\end{itemize} 
To compute $\hat\sigma_j$, we first find the variance $\sigma^2_j$ of 
$\hat \mu_j - \hat \theta_j \cdot b$. Using standard formulas for 
variances of sums, we have  
\begin{align} 
\sigma^2_j =\Var{ \hat \mu_j - \hat \theta_j \cdot b } &  =   
      \Var{ \bar y_j  - {\bar x}_j \cdot b }     \nonumber \\ 
& =  \Var{ \bar y_j } + \sum_{k=1}^s b_k^2 \Var{\bar x_j^k}.  
\label{eqn:s2j}  
\end{align} 
Under the multiplicative model 
the variance of $\bar y_j$ for a sample size of $n$ 
can be computed from Equation \ref{eqn:mvrel} as follows:
\begin{equation}
\Var{\bar y_j}  = \Var{ y_{i,j} }/n = 
        \mu_j^2 ( (\sigma^2_a +1 )(\sigma_e^2+1) -1 )/n. 
\label{eqn:vy} 
\end{equation}
Similarly, 
\begin{equation}
\Var{\bar x_j^k}  = \Var{ x_{i,j}^k }/n = 
        (\theta_j^k)^2 ( (\sigma^2_a +1 )( (\sigma_e^k)^2+1) -1 )/n. 
\label{eqn:vx} 
\end{equation}
where here $\sigma_e^1,\ldots, \sigma_e^s$ refer to the 
endmember-specific
measurement 
standard deviations described in Section 3.2. 
An estimate $\hat \sigma^2_j$  of $\sigma^2_j$ is then obtained by  
first 
replacing 
in Equations \ref{eqn:vy} and \ref{eqn:vx} 
the unknown values of $\mu_j$, $\theta_j^1,\ldots, \theta_j^s$, 
$\sigma^2_a$ and the $\sigma_e^2$'s with the estimates described in Section 3. 
The results are then plugged into the formula in Equation \ref{eqn:s2j} 
to obtain the estimate $\hat \sigma^2_j$ of $\sigma^2_j$. 
The denominator $\hat\sigma_j$ of the statistic $z_j$ is the square-root 
of $\hat\sigma_j^2$. 
Numerical examples of this calculation can be found in the replication 
material for this article. 

\end{document}